\documentclass[twocolumn,showpacs,preprintnumbers,superscriptaddress,floatfix,amsmath,amssymb]{revtex4}
\usepackage{amsfonts}
\usepackage{amsmath}
\usepackage{amssymb}
\usepackage{graphicx}%
\begin{document}
\title{Towards experimentally testing the paradox of black hole information loss}
\author{Baocheng Zhang}
\affiliation{State Key Laboratory of Magnetic Resonances and Atomic and Molecular Physics,
Wuhan Institute of Physics and Mathematics, Chinese Academy of Sciences, Wuhan
430071, China}
\author{Qing-yu Cai}
\email{qycai@wipm.ac.cn}
\affiliation{State Key Laboratory of Magnetic Resonances and Atomic and Molecular Physics,
Wuhan Institute of Physics and Mathematics, Chinese Academy of Sciences, Wuhan
430071, China}
\author{Ming-sheng Zhan}
\affiliation{State Key Laboratory of Magnetic Resonances and Atomic and Molecular Physics,
Wuhan Institute of Physics and Mathematics, Chinese Academy of Sciences, Wuhan
430071, China}
\affiliation{Center for Cold Atom Physics, Chinese Academy of Sciences, Wuhan 430071, China}
\author{Li You}
\email{lyou@mail.tsinghua.edu.cn}
\affiliation{State Key Laboratory of Low Dimensional Quantum Physics, 
Department of Physics, Tsinghua University, Beijing 100084, China}

\begin{abstract}
Information about the collapsed matter in a black hole will be lost if Hawking
radiations are truly thermal. Recent studies discover that information can be
transmitted from a black hole by Hawking radiations, due to their spectrum
deviating from exact thermality when back reaction is considered. In this
paper, we focus on the spectroscopic features of Hawking radiation from a
Schwarzschild black hole, contrasting the differences between the nonthermal
and thermal spectra. Of great interest, we find that the energy covariances of
Hawking radiations for the thermal spectrum are exactly zero, while the energy
covariances are non-trivial for the nonthermal spectrum. Consequently, the
nonthermal spectrum can be distinguished from the thermal one by counting the
energy covariances of successive emissions, which provides an avenue towards
experimentally testing the long-standing ``information loss paradox".

\end{abstract}

\pacs{04.70.Dy, 03.67.-a}
\maketitle

\section{Introduction}

Although not yet confirmed experimentally, it is widely believed that a black
hole evaporates almost like an ideal black body. This evaporation gives rise
to the famous Hawking radiation \cite{swh74,swh75}, where a thermal spectrum
\cite{rmw76,dnp76,wgu76} leads to a crisis of quantum physics
\cite{swh76,ls06}, the so-called \textquotedblleft information loss paradox".
In this scenario, information is lost as a black hole evaporates, inconsistent
with the unitarity of quantum mechanics, which thus presents a serious
obstacle for developing theories of quantum gravity. Many solutions aimed at
reconciling a thermal Hawking radiation spectrum with unitarity were discussed
\cite{acn87,kw89,jdb93,hm04,swh05,bt09}, none is capable of successfully
ending the dispute. As was pointed out by Wilczek and his collaborators
\cite{kw95,pw00}, the thermal spectrum is due to the neglect of back reaction
from emitted radiation in Hawking's original treatment. An improved derivation
even at a semiclassical level gives a spectrum slightly deviates from the
thermal one when energy conservation is enforced during the emission process
\cite{kw95,pw00,vad11}.

With the nonthermal spectrum of Parikh and Wilczek, we recently show that
Hawking radiations can carry off all information about the collapsed matter in
a black hole \cite{zcyz091,zcyz092}. After revealing the existence of
information-carrying correlation among Hawking radiations when the spectrum is
nonthermal, we show that entropy is conserved for Hawking radiation based on
standard probability theory and statistics. This prompts us to claim that the
information previously considered lost is hidden inside Hawking radiation, or
encoded into correlations. Our study thus establishes the basis for a
significant way towards resolving the long-standing information loss paradox
\cite{iy10}.

After discovering that the nonthermal spectrum of Parikh and Wilczek allows
for the emissions to carry off all information of a black hole
\cite{zcyz091,zcyz092,iy10}, a natural question arises as to whether Hawking
radiation is indeed nonthermal or not? Although the derivation of the
nonthermal spectrum is based on solid theoretical background, it remains to be
confirmed experimentally or in observations. In this paper, we explore
experimental signatures associated with a nonthermal spectrum of Hawking
radiation. Of great interest, we establish smoking gun like signatures capable
of distinguishing the slightly deviated nonthermal spectrum from the thermal
spectrum, based on an extensive analysis on the differences between the two
spectra. We thus proclaim that the information-carrying correlations among
Hawking radiations can be confirmed by observing the energy covariances of
successive emissions. Finally, we analyze the possibility of testing
experimentally the paradox by counting Hawking radiations.

\section{Uncovering the lost information}

We start by briefly reviewing our resolution to the paradox
\cite{zcyz091,zcyz092}. The nonthermal spectrum of Parikh and Wilczek
\cite{pw00} for a Schwarzschild black hole with mass $\mathcal{M}$ is given
by
\begin{equation}
\Gamma_{\mathrm{NT}}=\exp\left[  -8\pi{\mathcal{E}}\left(  {\mathcal{M}}%
-\frac{\mathcal{E}}{2}\right)  \right]  .
\end{equation}
From our earlier analysis \cite{zcyz091}, we use $S_{\mathrm{NT}}\left(
\mathcal{E}_{i}|\mathcal{E}_{1},\mathcal{E}_{2},\cdots,\mathcal{E}%
_{i-1}\right)  $ to denote the entropy of a Hawking emission, at an energy
$\mathcal{E}_{i}$, conditional on the earlier emissions labeled by
$\mathcal{E}_{1},\mathcal{E}_{2},\cdots$, and $\mathcal{E}_{i-1}$. As shown
before, it is easy to compute that
\begin{equation}
S_{\mathrm{NT}}\left(  \mathcal{E}_{i}|\mathcal{E}_{1},\mathcal{E}_{2}%
,\cdots,\mathcal{E}_{i-1}\right)  =8\pi\mathcal{E}_{i}\left(  \mathcal{M}%
-\sum_{j=1}^{i-1}\mathcal{E}_{j}-\mathcal{E}_{i}/2\right)  .
\end{equation}
Thus the total entropy for a given sequence of emissions $(\mathcal{E}%
_{1},\mathcal{E}_{2},\cdots,\mathcal{E}_{n})$ that exhaust the initial black
hole ($\mathcal{M}=\sum_{i=1}^{n}\mathcal{E}_{i}$) is given by
\begin{align}
\mathcal{S}_{\mathrm{NT}}\left(  \mathcal{E}_{1},\mathcal{E}_{2}%
,\cdots,\mathcal{E}_{n}\right)   &  =\sum_{i=1}^{n}S_{\mathrm{NT}}\left(
\mathcal{E}_{i}|\mathcal{E}_{1},\mathcal{E}_{2},\cdots,\mathcal{E}%
_{i-1}\right) \nonumber\\
&  =4\pi\mathcal{M}^{2}\nonumber\\
&  =S_{\mathrm{BH}},
\end{align}
where $S_{\mathrm{BH}}$ is the celebrated Bekenstein-Hawking entropy for a
black hole \cite{jdb73}. This equality states the entropy of all emitted
Hawking radiation is equal to the entropy of the initial black hole, which
implies no information is lost in the process of Hawking radiation
\cite{zcyz091,zcyz092,iy10}.

\section{Correlation among Hawking radiations}

With the nonthermal spectrum, correlations among Hawking radiations are
calibrated as follows: first we calculate the correlation between the emissions
of energies $\mathcal{E}_{1}$ and $\mathcal{E}_{2}$ and the result is
$8\pi\mathcal{E}_{1}\mathcal{E}_{2}$ \cite{zcyz091,zcyz092}; we then calculate
the correlation between the emissions with energies $\mathcal{E}%
_{1}+\mathcal{E}_{2}$ and $\mathcal{E}_{3}$ and the result becomes
$8\pi\left(  \mathcal{E}_{1}+\mathcal{E}_{2}\right)  \mathcal{E}_{3}$; the
total correlation among the three emissions $\mathcal{E}_{1}$, $\mathcal{E}%
_{2}$, and $\mathcal{E}_{3}$, thus becomes $8\pi\mathcal{E}_{1}\mathcal{E}%
_{2}+8\pi\left(  \mathcal{E}_{1}+\mathcal{E}_{2}\right)  \mathcal{E}_{3}$,
which does not depend on the orders of the individual emissions. For the
nonthermal spectrum, the total correlation in a queue of Hawking radiations is
summed up to
\begin{equation}
\mathcal{C}_{\mathrm{NT}}=\sum_{l\geq2}8\pi\left(  \mathcal{E}_{1}%
+\mathcal{E}_{2}+\cdots\mathcal{E}_{l-1}\right)  \mathcal{E}_{l}=\sum
_{i>j}8\pi\mathcal{E}_{i}\mathcal{E}_{j}, \label{nonthermal:correlation}%
\end{equation}
where indices $i,j$ take all possible values labeling emissions. Clearly, the
analogous correlation vanishes for a thermal spectrum, or $\mathcal{C}%
_{\mathrm{T}}=0$ \cite{swh76}. This will be discussed later as it plays a
crucial role in experimentally resolving the information loss paradox.

\begin{figure}[ptb]
\centering
\includegraphics[width=3.25in]{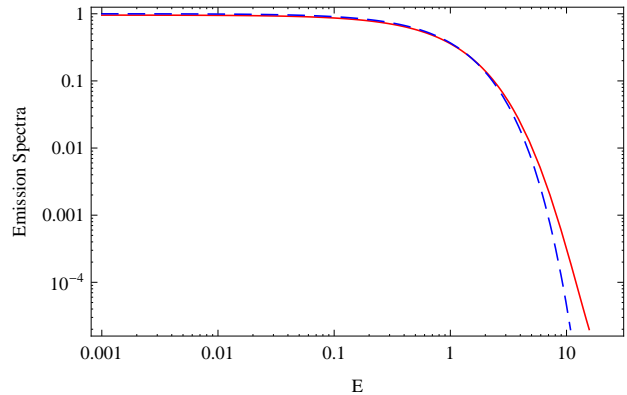}\caption{(Color online) The two spectra
compared for a black hole at the Planck mass scale. The red solid (blue
dashed) line refers to the nonthermal (thermal) spectrum. }%
\label{f1}%
\end{figure}

\section{Two spectra features compared}

Given a queue of emissions, we can compute the average energies and the
covariances of the emitted radiations for the two spectra: thermal or
nonthermal. These are studied below after we introduce suitable units for
discussing the physics of Hawking radiations and their associated properties.
Normal treatment of Hawking radiation assumes a static thermal black body
emission. Because the emission corresponds to a slow evaporation process for
most black holes, we can include the time dependence adiabatically. Using the
Stefan-Boltzmann's law for black body thermal emission and assuming no other
mass accretion or evaporation mechanisms exist, we can relate the reduction of
the mass for a black hole to the energies of the emissions according to
\begin{equation}
\frac{d\mathcal{M}}{dt}=-4\pi R^{2}\sigma T^{4},
\end{equation}
where $R=2\mathcal{M}$ is the Schwarzschild radius, the temperature of Hawking
radiation is given by $T={1}/{8\pi}\mathcal{M}$, inversely proportional to its
mass, and the Stefan-Boltzmann constant $\sigma={2\pi^{5}k_{b}^{4}}%
/({15c^{2}h^{3}})$. This gives the steady state time dependent mass
\begin{equation}
\mathcal{M}\left(  t\right)  =\mathcal{M}_{0}-\left(  \frac{3\sigma}%
{256\pi^{3}}\right)  ^{\frac{1}{3}}\left(  t-t_{0}\right)  ^{\frac{1}{3}},
\label{time:dependent:mass}%
\end{equation}
of a black hole if the total mass change is due to Hawking radiation alone. In
the above, $\mathcal{M}_{0}$ denotes the initial mass of a black hole at time
$t_{0}$. In subsequent emissions, the mass thus becomes a time unit according
to Eq. (\ref{time:dependent:mass}). Therefore, some of the figures illustrated
below are plotted against mass instead of time.

To prepare for numerically comparison of the observables associated with the
two spectra, we normalize both according to $\int_{0}^{\mathcal{M}}%
\Gamma(\mathcal{E})d\mathcal{E}=1$ and obtain respectively the normalized
thermal spectrum
\begin{equation}
\Gamma_{\mathrm{T}}\left(  \mathcal{E}\right)  =8\pi\mathcal{M}^{2}\exp\left(
-8\pi\mathcal{M}\mathcal{E}\right)  /[1-\exp\left(  -8\pi\mathcal{M}%
^{2}\right)  ],
\end{equation}
and the nonthermal spectrum
\begin{equation}
\Gamma_{\mathrm{NT}}\left(  \mathcal{E}\right)  =2\sqrt{\pi}\mathcal{M}%
\exp\left(  -8\pi\mathcal{E}\left(  M-\mathcal{E}/2\right)  \right)
/\operatorname{F}(2\sqrt{\pi}\mathcal{M}),
\end{equation}
where $\mathrm{F}(x)$ is the Dawson function \cite{as72}. We can now discuss
the relevant features of the two spectra in quantitative terms. We will
illustrate their differences graphically. As a result of quantum mechanics
meeting with gravity, the spectrum for Hawking radiation is a function of
physical quantities from very large ($c$) to very small ($\hbar$ and $G$).
Therefore, it is more convenient for numerical purposes to adopt real units
instead of the commonly used units of $G=c=\hbar=1$.

\begin{figure}[ptb]
\centering
\includegraphics[width=3.25in]{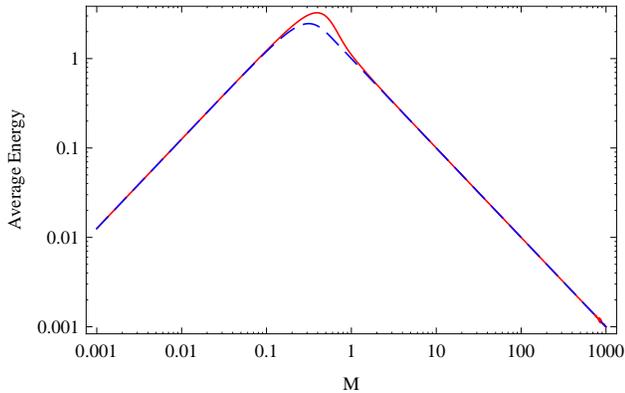}\caption{(Color online) The average
emission energies compared, as in Fig. \ref{f1}.}%
\label{f2}%
\end{figure}

Introducing the dimensionless mass $M=\mathcal{M}/\mathcal{M}_{\mathrm{pl}}$
and energy $E=\mathcal{E}/k_{b}T_{\mathrm{pl}}$ in terms of Planck mass
${\mathcal{M}_{\mathrm{pl}}}$ and Planck temperature $T_{\mathrm{pl}}$, we
arrive at the respectively normalized spectra
\begin{equation}
\Gamma_{\mathrm{NT}}\left(  E\right)  =\exp\left[  -\left(  M-E/{16\pi
}\right)  E\right]  /4\sqrt{\pi}\mathrm{F}(2\sqrt{\pi}M),
\end{equation}
and
\begin{equation}
\Gamma_{\mathrm{T}}\left(  E\right)  =M\exp\left(  -ME\right)  /[1-\exp\left(
-8\pi M^{2}\right)  ],
\end{equation}
with $E\in\lbrack0,8\pi M]$. Figure \ref{f1} compares the two spectra for a
black hole at Planck scale. Their difference is concentrated near $E\sim
k_{b}T_{M}$, where $T_{M}$ denotes the equivalent Hawking radiation
temperature for a black hole of mass $\mathcal{M}$ measured in units of
$T_{\mathrm{pl}}$.

In units of $k_{b}T_{\mathrm{pl}}$, the average energy of Hawking emissions
can be computed at any instant, as for a fixed mass black hole. For the
thermal spectrum, we find
\begin{equation}
\label{thermal:emission:energy}\langle E(M)\rangle_{\mathrm{T}}=\left(
M{{{+{\frac{8\pi M^{3}}{{e^{8\pi M^{2}}-1-8\pi M^{2}}}}}}}\right)  ^{-1},
\end{equation}
which approaches ${4\pi}M$ for $M\ll1$, and $1/M$ for $M\gg1$. This is to be
compared with the nonthermal spectrum, for which we find
\begin{equation}
\label{nonthermal:emission:energy}\langle E(M)\rangle_{\mathrm{NT}}=8\pi
M-\frac{2\sqrt{\pi}(1-e^{-4\pi M^{2}})}{\mathrm{F}(2\sqrt{\pi}M)}.
\end{equation}
Interestingly both limits are the same as for the thermal spectrum discussed
above for $M\ll1$ and $M\gg1$. Their noticeable difference exists only near
the Planck scale as shown in Fig. \ref{f2}.

\begin{figure}[ptb]
\centering
\includegraphics[width=3.25in]{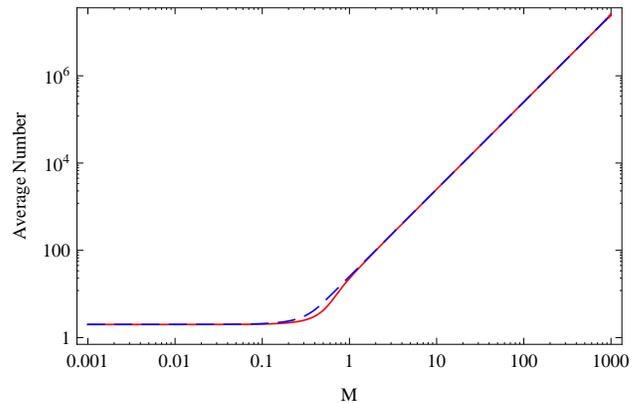}\caption{(Color online) The average
number of emissions compared for the two spectra as in Fig. \ref{f1}.}%
\label{f3}%
\end{figure}

The average number of radiations emitted from a black hole with mass $M$ can
be obtained for the thermal spectrum approximately as
\begin{equation}
\label{thermal:averaged:number}N_{\mathrm{T}}(M)=\frac{{8\pi}M}{\langle
E(M)\rangle_{\mathrm{T}} }=\frac{{8\pi}\left(  {e^{8\pi M^{2}}-1-8\pi M^{2}%
}\right)  }{{e^{8\pi M^{2} }-1}},
\end{equation}
while for the nonthermal spectrum we find
\begin{equation}
\label{nonthermal:averaged:number}N_{\mathrm{NT}}(M)=\frac{{4\pi}%
M\mathrm{F}(2\sqrt{\pi}M)}{{4\pi}M\mathrm{F}(2\sqrt{\pi}M)-\sqrt{\pi
}(1-e^{-4\pi M^{2}})}.
\end{equation}
Again both Eqs. (\ref{thermal:averaged:number})
(\ref{nonthermal:averaged:number}) give the same limits: ${2}$ for $M\ll1$,
and $8{\pi}M^{2}$ for $M\gg1$. As shown in Fig. \ref{f3}, the average number
of emissions increase rapidly with the black hole mass. Below the Planck
scale, however, the average number of emissions remains nearly a constant.

\begin{figure}[ptb]
\centering
\includegraphics[width=3.25in]{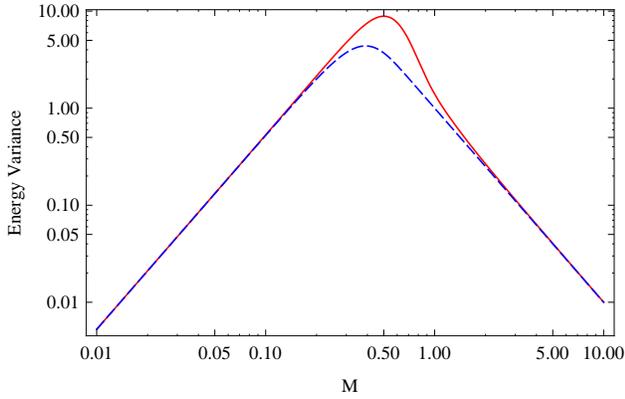}\caption{(Color online) The radiation
energy variances compared for the two spectra as in Fig. \ref{f1}. }%
\label{f4}%
\end{figure}

We now compute the standard deviations of the emission energies, and we find
for the thermal spectrum
\begin{align}
\delta E_{\mathrm{T}}^{2}(M)  &  =\langle E^{2}(M)\rangle_{\mathrm{T}}-\langle
E(M)\rangle_{\mathrm{T}}^{2}\nonumber\\
&  =\frac{(\cosh{8\pi M^{2}}-1-32\pi^{2}M^{4})\mathrm{csch}^{2}{4\pi M^{2}}%
}{2M^{2}},\hskip18pt \label{thermal:standard:deviations}%
\end{align}
whose large and small $M$ limits are given respectively by ${16\pi^{2}M}%
^{2}/3$ for $M\ll1$, and $1/M^{2}$ for $M\gg1$. For the nonthermal spectrum,
we find
\begin{align}
\delta E_{\mathrm{NT}}^{2}(M)  &  ={\frac{4\pi}{{\mathrm{F}^{2}({2}\sqrt{{\pi
}}M)}}}\Big[4\sqrt{\pi}M\mathrm{F}({2}\sqrt{{\pi}}M)\nonumber\\
&  \hskip64pt -(1-e^{-4\pi M^{2}})^{2}\Big]-8\pi, \hskip 18pt
\label{nonthermal:standard:deviations}%
\end{align}
again with the same large and small $M$ limits as for the thermal spectrum.
Figure \ref{f4} shows clearly these features and illustrates the dependence of
the variances on the average energy.

Based on our extensive analysis, we find no drastic differences between the
two spectra except for tiny black holes with masses near the Planck scale.
Naively, one would conclude that it is therefore essentially impossible to
experimentally distinguish the nonthermal spectrum from the thermal one, let
alone to query about whether information can be carried out from a black hole
by correlations hidden in the emissions. Nevertheless, we demonstrate below
that information stored in the correlations of Hawking radiations from the
nonthermal spectrum can indeed be observed through a counting of the emission
energy covariances.

\section{Energy covariances}

\begin{figure}[ptb]
\centering
\includegraphics[width=3.25in]{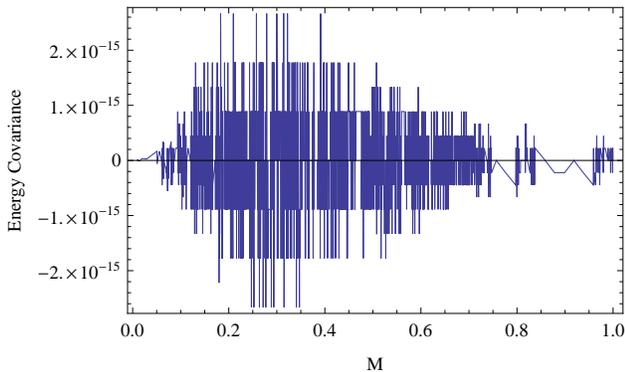}\caption{(Color online) The covariances
of successive emissions are zero for the thermal spectrum, \textit{i.e.},
there is no correlation among Hawking radiations for the thermal spectrum. The
``zero'' noise feature is a result of numerical round-off error.}%
\label{f5}%
\end{figure}

For the thermal spectrum, individual emissions are uncorrelated \cite{swh76},
thus one expects a vanishing covariance. This is indeed what we find
\begin{align}
\delta E_{\mathrm{T}}^{2\mathrm{(cov)}}  &  =\langle E_{i}(M)E_{j\neq
i}(M)\rangle_{\mathrm{T}}-\langle E_{i}(M)\rangle_{\mathrm{T}}\langle E_{j\neq
i}(M)\rangle_{\mathrm{T}}\nonumber\label{thermal:covariance}\\
&  =0.
\end{align}
In confirming the above result, $\langle E_{i}(M)\rangle_{\mathrm{T}}=\langle
E_{j\neq i}(M)\rangle_{\mathrm{T}}=\langle E(M)\rangle_{\mathrm{T}}$ is
obtained when individual emission energies are averaged over an ideal black
body spectrum. This is shown in Fig. \ref{f5}, where as we have stated loud and clear
repeatedly no correlation among different thermal emissions exists.

For the nonthermal spectrum, the average cross energy term $\langle
E_{i}(M)E_{j\neq i}(M)\rangle$ is nontrivial because of the existence of
correlation. It is governed by the probability for two emissions: one at an
energy $E_{i}$ and another at an energy $E_{j}$. This probability was derived
by us in Ref. \cite{zcyz092}, where we find that for an extensive list of
black holes it satisfies $\Gamma_{\mathrm{NT}}(E_{1},E_{2})=\Gamma
_{\mathrm{NT}}(E_{1}+E_{2})$. In fact, a recursive use of this relation allows
us to show
\begin{align}
\Gamma_{\mathrm{NT}}(E_{1},E_{2})  &  =\Gamma_{\mathrm{NT}}(E_{1}+E_{2}
)\nonumber\\
&  =\Gamma_{\mathrm{NT}}(E_{1}^{\prime},E_{1}+E_{2}-E_{1}^{\prime}),\hskip12pt
\label{nonthermal:correlation:relation}%
\end{align}
as long as $E_{1}+E_{2}-E_{1}^{\prime}>0$, or the probability for emissions
$E_{1},E_{2},E_{3},\cdots$ is the same as the probability for the emission of
a single radiation with an energy $\sum_{j}E_{j}$. This probability
distribution is symmetric with respect to any permutation of the individual
emission indices. Thus we can work within one sector, and define the
normalized probability subjected to the energy conservation constraint
$\sum_{j}E_{j}\in\lbrack0, 8\pi M]$.

The spectrum for multiple emissions thus takes the form,
\begin{equation}
\Gamma_{\mathrm{NT}}(\sum_{j}E_{j})\sim{\exp\left[  -(M-\sum_{j}E_{j}/{16\pi
})\sum_{j}E_{j}\right]  },
\end{equation}
which is symmetric with respect to all permutations of indices. $\Gamma
_{\mathrm{NT}}(E_{1},E_{2},E_{3}\cdots)$ can be normalized according to
$\int_{0}^{8\pi M}dE_{1}\int_{0}^{8\pi M\mathcal{-}E_{1}}dE_{2}\cdots
\Gamma_{\mathrm{NT}}(E_{1},E_{2},E_{3}\cdots)=1$. For the case of two
emissions, the explicit form is found to be $\Gamma_{\mathrm{NT}}(E_{1}%
,E_{2})=\exp\{-[M-{(E_{1}+E_{2})}/({16\pi})](E_{1}+E_{2})\}/8\pi
\lbrack-1+e^{-4\pi M^{2}}+4\sqrt{\pi}M\mathrm{F}(2\sqrt{\pi}M)],$which gives
\begin{widetext}
\begin{eqnarray}
\langle E_{i}(M)E_{j\neq i}(M)\rangle
=\frac{8\pi\left[  \left(  12\pi M^{2}-1\right)e^{-4\pi
M^{2}}+1-4\pi M^{2}+2\sqrt{\pi}M\left(8\pi M^{2}-3\right)
\mathrm{F}(2\sqrt{\pi}M)\right]}{3[-1+e^{-4\pi
M^{2}}+4\sqrt{\pi}M\mathrm{F}(2\sqrt{\pi}M)]}.
\label{averaged:correlation}
\end{eqnarray}
Together with the result of Eq. (\ref{nonthermal:emission:energy}) for
$\langle E(M)\rangle$, we find
\begin{eqnarray}
\delta E_{\mathrm{NT}}^{2\mathrm{(cov)}}(M)
&=&{\frac{2\pi}{3\mathrm{F}^{2}(2\sqrt{\pi}M)\left[  e^{-4\pi M^{2}}%
-1+4\sqrt{\pi}M\mathrm{F}(2\sqrt{\pi}M)\right]  }}\nonumber\\
&  &
\left[  72\sqrt{\pi}M\left(
1-{e^{-4\pi M^{2}}}\right)  ^{2}\mathrm{F}\left(  2\sqrt{\pi}M\right)
\right.  -4\left(  1+60\pi M^{2}\right)  {e^{-4\pi M^{2}}\mathrm{F}^{2}}(2\sqrt{\pi
}M)\nonumber\\
&  & +6\left(  1-{e^{-4\pi M^{2}}}\right)^{3}+4\left(  1+68\pi
M^{2}\right) {\mathrm{F}^{2}}(2\sqrt{\pi}M) {-8\sqrt{\pi}}M\left(
3+40\pi M^{2}\right) {\mathrm{F}^{3}}(2\sqrt{\pi }M)\bigg].
\label{covariance}
\end{eqnarray}
\end{widetext}The two limits are given respectively by
\begin{align}
\delta E_{\mathrm{NT}}^{2\mathrm{(cov)}}(M\rightarrow0)  &  \sim-{\frac
{32\pi^{2}M^{2}}{3}}+{\frac{96\pi^{3}M^{4}}{5}}+\cdots,\hskip18pt
\label{small}\\
\delta E_{\mathrm{NT}}^{2\mathrm{(cov)}}(M\rightarrow\infty)  &  \sim
-{\frac{29}{16\pi M^{4}}}. \label{large}%
\end{align}

\begin{figure}[ptb]
\centering
\includegraphics[width=3.25in]{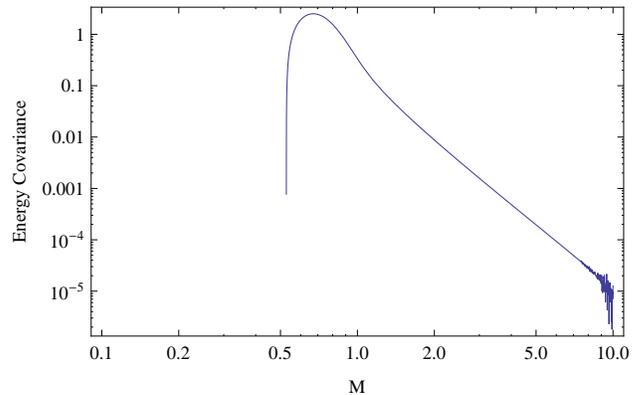}\caption{(Color online) The covariances
of successive emissions is non-trivial for the nonthermal spectrum. One can
distinguish the nonthermal spectrum from the thermal one by counting the
covariances. This provides an avenue towards experimentally testing the
paradox.}%
\label{f6}%
\end{figure}

Upon checking for the quantitative results of Fig. \ref{f6}, we find that the
covariance approaches its maximum also near the Planck scale. It is
interesting that the covariance vanishes at small or large masses. As is
discussed in Fig. \ref{f3}, the number of emissions becomes limited (for
instance, 2 emissions) when its mass is small. The covariance thus vanishes at
the small mass limit [$M\rightarrow0$ in Eq. (\ref{small})]. For large masses,
the curve of covariance decreases because the average emission energy becomes
small, as illustrated in Fig. \ref{f2}. The correlation between two emissions
$E_{i}$ and $E_{j}$ is proportional to the product of $E_{i}$ and $E_{j}$, so
it is reasonable that the covariance decreases at large mass limit [$M
\rightarrow\infty$ in Eq. (\ref{large})].

\section{Discussion and conclusion}

In reality, the effective radiation temperature for any astrophysical black
hole is much less than the cosmic background temperature. So, the possibility
of the Hawking radiation observation focuses on the micro black holes which
have been discussed extensively in connection with the experiments of the CERN
Large Hadron Collider (LHC) \cite{ehm00,dl01,gt02,mr08,cms11}. In this
regards, the D-dimensional Schwarzschild metric is required and the
fundamental Planck scale is reduced depending on the compact space of volume
$V_{D-4}$, e.g. the reduced Planck scale $M_{pl}\sim1Tev$ with $D=10$ and
$V_{6}\sim fm^{6}$ \cite{gt02}. A recent estimate showed that the minimum
black hole mass should be in the range of $3.5-4.5$ TeV for $pp$ collisions at a
center-of-mass energy of $7$ TeV at LHC \cite{cms11}. According to our
analysis, the energy covariance (\ref{covariance}) for the nonthermal spectrum
is larger near the Planck scale, i.e. $\delta E_{\mathrm{NT}}^{2\mathrm{(cov)}%
}$ $\sim1$ for a black hole mass $\sim0.7M_{pl}$, which shows 
that if the radiation of a micro black hole were observed at LHC, 
its energy covariance will determine whether the emission spectrum 
is indeed nonthermal. Of course, the energy
scale about the production and observation of micro black holes is being
debated \cite{mns12} and so it remains to be seen when and whether the micro black holes
could be observed on Earth, especially in LHC experiment.

On the other hand, observation of Hawking radiation at small
man-made black holes such as those implemented or discussed with optical,
acoustic, and cold-atomic systems \cite{pkr08,bcc10,u81,rmm08,wtp11,lbg10} 
are being discussed, and several experiments \cite{pkr08,bcc10,rmm08,wtp11,lbg10} had
shown the evidence of the Hawking radiation from the event horizon. For
instance, the observation of the analogous
Hawking radiation from ultrashort laser pulse filaments was recently reported 
by creating a
traveling refractive index perturbation (RIP) in the fused silica glass
\cite{bcc10}. In the experiment, the analogous temperature of Hawking
radiation is $T_{H}\sim10^{-3}K$ which is equivalent to the radiation of a
black hole with mass $M\sim10^{26}kg$. Such mass is large enough to make the
average energy covariance is ignorable, i.e. $\delta E_{\mathrm{NT}%
}^{2\mathrm{(cov)}}\sim10^{-120}$. Actually, the nonthermal probability in
real units could be written as $\Gamma_{NT}=\exp\left[  -8\pi E\left(
M-\frac{E}{2c^{2}}\right)  \frac{G}{\hbar c^{3}}\right]  =\exp\left[  -8\pi
EM\frac{G}{\hbar c^{3}}\right]  \exp\left[  8\pi E^{2}\frac{G}{2\hbar c^{5}%
}\right]  =\Gamma_{T}\Delta\Gamma$ where $\Delta\Gamma=\exp\left[  -8\pi
E^{2}\frac{G}{2\hbar c^{5}}\right]  $ is the correction factor. If we want to
observe the nonthermal spectrum, it is important to make $\Delta\Gamma$
deviate significantly from 1. For the temperature $T_{H}\sim10^{-3}K$, 
the most probable
emitted energy is $E\sim10^{-25}J$, which shows $\Delta\Gamma\sim1$ and so
it is impossible to distinguish the nonthermal and thermal spectra, 
in this case unless emissions with significantly larger E can be observed. 
It is estimated that the
observable $\Delta\Gamma$ requires the emissions $E$ with ultra-high energy,
e.g., $\Delta\Gamma\sim\Gamma_{T}$ indicates $E\gg1J$
which cannot happen in the discussed physical system.
It is noticed that the analysis here
bases on the correspondence between optical black hole and Schwarzschild black
hole, so it remains to be seen whether this correspondence is permitted when
the backreaction is included in the radiation process and it is still
interesting to see whether the correlation between emissions of Hawking
radiation could be observed since the unitarity requires the existence of such
correlations \cite{zcyz091,zcyz092,iy10}.

In summary, the important differences between the thermal and the nonthermal
spectra, such as their respective average emission energies, average numbers
of emissions, average emission energy fluctuations, and energy covariances,
are calculated, compared, and discussed. Based on our earlier result that
so-called lost information of a black hole is encoded into correlations among
Hawking radiations, we can verify whether the spectrum of Hawking radiation is
nonthermal or not by counting the energy covariances of Hawking radiations,
which presents an avenue towards experimentally testing the paradox of black
hole information loss.

\section{\bigskip Acknowledgement}

Financial support from NSFC under Grant Nos. 11074283, 11104324 and 91121005,
and NBRPC under Grant Nos.2010CB832805, No. 2012CB922100, 
and No. 2013CB922000 is gratefully
acknowledged.

\end{document}